\documentclass[twocolumn,twocolappendix]{aastex631}
\usepackage{graphicx}
\usepackage{xcolor,graphicx,xspace,color,longtable}
\usepackage{amssymb,amsmath,shadow,bezier,curves,rotating}
\usepackage{graphicx,chngcntr}
 \usepackage{booktabs}
\graphicspath{ {./images/}}

\newcommand {\h}    {$h^{-1}\,\mathrm{Mpc}$}

\newcommand {\hm}   {$h^{-1} \  M_{\odot}$}
\newcommand {\hms}   {$h^{-2} \  M_{\odot}$}
\newcommand {\m}    {$M_{\odot}$}

\usepackage{xcolor}
\begin{document}

\title[BSG Stellar Mass-Halo Mass Relation] {
Hidden Connections: Tracing the BCG Stellar Mass–Halo Mass Relation in the SDSS–GalWCat19 Cluster Catalog}

\author[0000-0003-3595-7147]{Mohamed H. Abdullah}
\affiliation{Department of Physics, University of California Merced, 5200 North Lake Road, Merced, CA 95343, USA}
\affiliation{Department of Astronomy, National Research Institute of Astronomy and Geophysics, Cairo, 11421, Egypt}

\author[0000-0003-2716-8332]{Rasha M. Samir}
\affiliation{Department of Astronomy, National Research Institute of Astronomy and Geophysics, Cairo, 11421, Egypt}

\author[0009-0003-6781-1895]{Nouran E. Abdelhamid}
\affiliation{Department of Astronomy, National Research Institute of Astronomy and Geophysics, Cairo, 11421, Egypt}

\author[0009-0008-3612-8942]{Shrouk Abdulshafy}
\affiliation{Department of Physics, University of California Merced, 5200 North Lake Road, Merced, CA 95343, USA}
\affiliation{Department of Astronomy, Cairo University, 1 Gamaa Street, Giza, 12613, Egypt}

\author[0000-0002-6572-7089]{Gillian Wilson}
\affiliation{Department of Physics, University of California Merced, 5200 North Lake Road, Merced, CA 95343, USA}

\author{Anatoly Klypin}
\affiliation{Astronomy Department, New Mexico State University, Las Cruces, NM 88001}
\affiliation{Department of Astronomy, University  of Virginia, Charlottesville, VA 22904, USA}

\author[0000-0003-3427-1733]{SH. M. Shehata}
\affiliation{Department of Astronomy, National Research Institute of Astronomy and Geophysics, Cairo, 11421, Egypt}
\affiliation{LUX, CNRS UMR 8262, Observatoire de Paris, 61 Avenue
de l'Observatoire, 75014 Paris, France}
\author[0000-0002-4216-1908]{Ashraf A. Shaker}
\affiliation{Department of Astronomy, National Research Institute of Astronomy and Geophysics, Cairo, 11421, Egypt}

\begin{abstract}
The stellar-to-halo mass (SMHM) relation of brightest cluster galaxies (BCGs) provides key insight into the connection between BCG growth and the assembly of their host halos.
We analyze this relation using the spectroscopic SDSS GalWCat19 cluster catalog, selecting 996 systems with $\log(M_{200})\ge 13.6$~[\hm], $\log(M_{\star})\ge 10.5$~[\hms], and $0.02 \le z \le 0.125$ to limit evolutionary effects and ensure stellar-mass completeness.
We fit lognormal scaling relations with a Markov Chain Monte Carlo (MCMC) framework that accounts for measurement uncertainties and intrinsic scatter.
For the fiducial SMHM relation, $\langle \log M_{\star}\,|\,M_{200}\rangle = \alpha + \beta \log(M_{200}/M_{\mathrm{piv}})$ with $\log(M_{\mathrm{piv}})=14.2$, we find a shallow slope $\beta = 0.17 \pm 0.03$, normalization $\alpha = 11.04 \pm 0.01$, and intrinsic scatter $\sigma_{\mathrm{int}} = 0.19 \pm 0.01~\mathrm{dex}$.
Recasting the relation in normalized form reduces the scatter to $0.16 \pm 0.01~\mathrm{dex}$, while including the magnitude gap $M_{14}$ further reduces it to $0.14 \pm 0.01~\mathrm{dex}$.
Variations in richness, redshift, and mass thresholds produce systematic shifts that are small compared to the statistical uncertainties, indicating that our inferred relations are robust to plausible selection choices.
The reduced scatter when including $M_{14}$ supports a picture in which BCG stellar mass reflects both halo mass and halo assembly history.
\end{abstract}

\section{Introduction}

Galaxy clusters are the most massive virialized systems in the Universe and serve as fundamental tracers of both large-scale structure formation and galaxy evolution. Their deep gravitational potential wells, dominated by dark matter, host hot intracluster gas and hundreds to thousands of galaxies. These components make clusters ideal laboratories for exploring the interplay between baryons and dark matter, and for probing the co-evolution of galaxies within their large-scale environments. Clusters also exhibit a range of empirical scaling relations, linking mass, richness, X-ray properties, and stellar content, that provide valuable insights into cluster assembly histories and are widely used to constrain cosmological parameters \citep{Voit05,Allen11,Kravtsov12}.

Among these relations, the stellar-to-halo mass (SMHM) relation, linking the stellar mass of the brightest cluster galaxy (BCG) to the total halo mass, is particularly important. It plays a critical role in understanding how baryonic and dark matter components co-evolve. BCGs are the most massive and luminous galaxies in the Universe, typically residing near the centers of galaxy clusters where they experience frequent interactions and mergers. Their formation is primarily driven by hierarchical merging, accretion of intracluster stars, and dynamical processes within the dense cluster environment \citep{DeLucia07,Lavoie16,Zhang16}. These features make BCGs ideal tracers of halo growth and central galaxy evolution.

BCGs are generally characterized as massive, elliptical galaxies with smooth light profiles and predominantly old stellar populations, often described as passively evolving systems \citep{Edwards2016,Kormendy2016}. They differ from normal ellipticals in several respects: BCGs exhibit distinct surface brightness profiles \citep{Bernardi2007,VonDerLinden2007,Lin2010,Tonnesen2021, Chu2021, Chu2022, 2022ApJ...933..215D, Ellien2025}, small peculiar velocities relative to other cluster galaxies, and in some cases signatures of residual star formation or AGN activity \citep{Donahue2010,Donahue2017,Fogarty2019,Calzadilla2022,Levitskiy2024,Saxena2024}. 
Recent deep imaging surveys with JWST and Hyper Suprime-Cam have illustrated that BCGs show multi-component surface brightness profiles extending to several hundred kpc, with clear structural transitions between the central galaxy and intracluster light components. This confirms that the hierarchical growth of BCGs is built through mergers and tidal stripping \citep{Montes2021, Dalal2021, Ellien2025, Ghodsi2025}.
These unique properties underscore their close connection to cluster-scale processes and the importance of linking BCG evolution to that of their host halos.

A number of studies have investigated the SMHM relation for BCGs, establishing its central role in linking luminous and dark matter in clusters. Early work by \citet{Gonzalez2013} demonstrated a tight correlation between BCG stellar mass and halo mass, suggesting that BCGs can serve as reliable tracers of cluster-scale dark matter halos. Subsequent efforts extended these results to larger samples and higher redshifts, revealing that the relation may evolve with cosmic time \citep[e.g.,][]{Golden-Marx19,Erfanianfar19, Golden-Marx2022, Chiu2025}. Studies using both observations and hydrodynamical simulations have likewise suggested redshift evolution, with BCGs undergoing rapid mass growth at early times followed by slower accretion at later epochs \citep{Lin2004, Kravtsov18, DeMaio2020, Gozaliasl2024}. In contrast, \citet{Erfanianfar19}, using a catalog of 416 BCGs out to $z \sim 0.65$, reported no significant evolution in the relation, highlighting continuing uncertainties in the dominant mechanisms of BCG growth.

Different observational approaches have reinforced the robustness of the SMHM relation. Weak-lensing studies using CFHTLenS \citep{Erben13} and DES \citep{DES16} data confirmed the correlation across wide halo mass ranges \citep{Leauthaud2012, 2016MNRAS.459.3251V}, while X-ray and dynamical analyses provided complementary constraints \citep{Simet2017, Kravtsov18, Erfanianfar19}. Multiwavelength studies have further highlighted environmental dependencies: correlations with cluster richness and X-ray luminosity \citep{Furnell2018, Miyaoka2018}, as well as links to galaxy morphology, structure, and star formation activity \citep{Zhao2015a,Zhao2015b,Gozaliasl2016}. The dynamical state of clusters has also been shown to influence the relation, with more relaxed systems hosting more massive BCGs at fixed halo mass \citep{2019MNRAS.484.2807L, Sohn2020, Pasini2022}. Consistently, \cite{Samir2021} found that BCG stellar properties correlate strongly with the dynamical parameter of the host, cluster velocity dispersion, further reinforcing the connection between BCG growth and cluster environment.

\begin{figure*}\hspace{-0cm}   \includegraphics[width=1\linewidth]{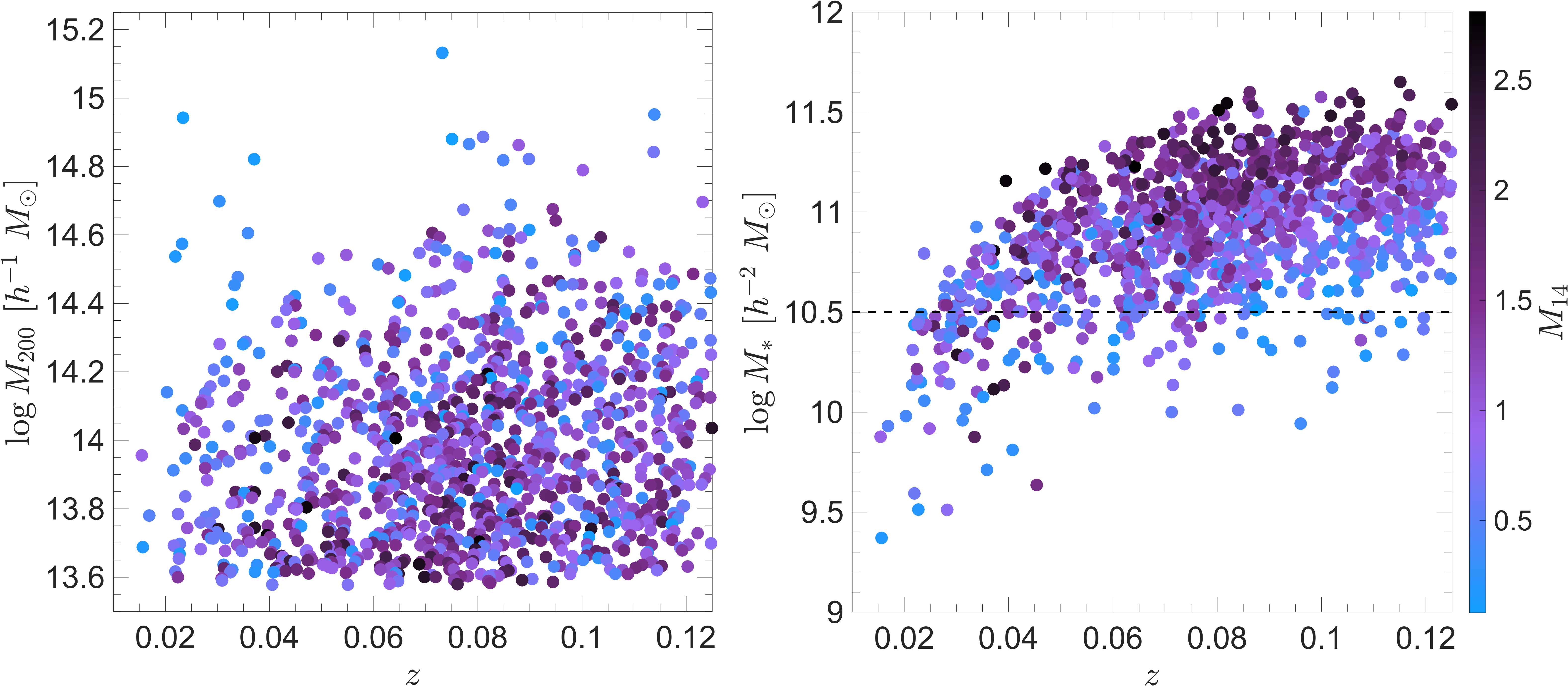} \vspace{-0.5cm}
    \caption{
    Left: Distribution of galaxy cluster masses $M_{200}$ as a function of redshift for our sample of 1185 clusters. The sample is approximately complete for clusters with $\log{M_{200}} \geq 13.6$ [\hm] in the redshift range $0.02 \leq z \leq 0.125$. 
    Right: Distribution of BCG stellar masses $\mathrm{M}_{\ast}$ versus redshift. To ensure completeness, we select BCGs with $\log{\mathrm{M}_{\ast}} \geq 10.5$ [\hms]. 
    In both panels, the color bar indicates the magnitude gap $M_{14}$, defined as the brightness difference between the BCG and the fourth brightest galaxy member.
    }   
\label{fig:redshift}
\end{figure*}
Despite extensive progress, the SMHM relation still exhibits non-negligible scatter, reflecting both the diversity of cluster assembly histories and methodological systematics. 

A promising secondary parameter is the magnitude gap between the brightest cluster galaxy and the fourth-brightest cluster member galaxy, $M_{14}\equiv M_4 - M_1$, where $M_1$ and $M_4$ are the $r$-band absolute magnitudes (k-corrected and evolution-corrected to $z=0.1$) of the brightest and fourth-brightest member galaxies within $0.5R_{200}$, which correlates with the cluster dynamical state: larger gaps are often linked to older, more relaxed systems and may encode information about cluster assembly \citep[e.g.,][]{Raouf14,Golden-Marx18}.
Incorporating $M_{14}$ has been shown to reduce the intrinsic scatter in the normalized SMHM relation \citep{Golden-Marx22}, highlighting its utility as a probe of BCG and cluster co-evolution. At the same time, discrepancies among previous studies arise from differences in cluster selection, membership determination, and stellar mass estimates \citep{Zhang16,Lin2017,Golden-Marx19}. Many analyses based on photometric catalogs (e.g., RedMAPPER; \citealt{Rykoff2014}) suffer from uncertainties in richness-based mass calibration, while spectroscopic surveys offer more robust dynamical mass estimates that help reduce scatter in the relation \citep{Muzzin2012, 2016MNRAS.461..248S}.

In this work, we utilize the spectroscopic \(\mathtt{GalWCat19}\)\footnote{\url{http://cdsarc.u-strasbg.fr/viz-bin/cat/J/ApJS/246/2}} galaxy cluster catalog \citep{Abdullah20a}, constructed from SDSS DR13 data \citep{Albareti17}. The catalog is based on the GalWeight technique \citep{Abdullah18}, which accounts for redshift-space distortions and avoids reliance on empirical assumptions or iterative procedures, thereby enabling accurate cluster membership determination and robust dynamical mass estimates. The low-redshift nature of \(\mathtt{GalWCat19}\) (\(z \lesssim 0.2\)) minimizes the need for evolutionary corrections, making it particularly well-suited for this analysis. Galaxy stellar masses (\(\mathrm{M}_\ast\)) are obtained from the MPA/JHU value-added catalog\footnote{\url{http://www.mpa-garching.mpg.de/SDSS/DR7/}}. 
Together, these datasets provide a reliable foundation to quantify the impact of selection systematics on the SMHM relation and to evaluate the role of $M_{14}$ in reducing its intrinsic scatter. Using a large sample of spectroscopically confirmed clusters, we present a comprehensive analysis of the $M_{\mathrm{BCG}} - M_{200}$ relations and compare our results with previous studies, offering new insights into the connection between BCG growth and cluster evolution and its implications for hierarchical structure formation and the physical processes shaping central galaxies.

This paper is organized as follows. In Section~\ref{sec:galw}, we describe the cluster sample and data selection. Section~\ref{sec:method} outlines our methodology for fitting the SMHM relations. In Section~\ref{sec:results}, we present the fitted SMHM relations, assess the intrinsic scatter and the role of \(M_{14}\), examine potential systematic effects, and compare our results with previous observations and models. Finally, we summarize our results and conclusions in Section~\ref{sec:conc}. Throughout this paper, we adopt a flat $\Lambda$CDM cosmology consistent with the \citet{Planck15} results, assuming $\Omega_{\mathrm{M}} = 0.3089$, $\Omega_{\Lambda} = 0.6911$, and $h = 0.6774$. We use the term `log' to refer to base-10 logarithms (i.e., $\log_{10}$) throughout.

\begin{figure*}\hspace{-0cm}
\includegraphics[width=1\linewidth]{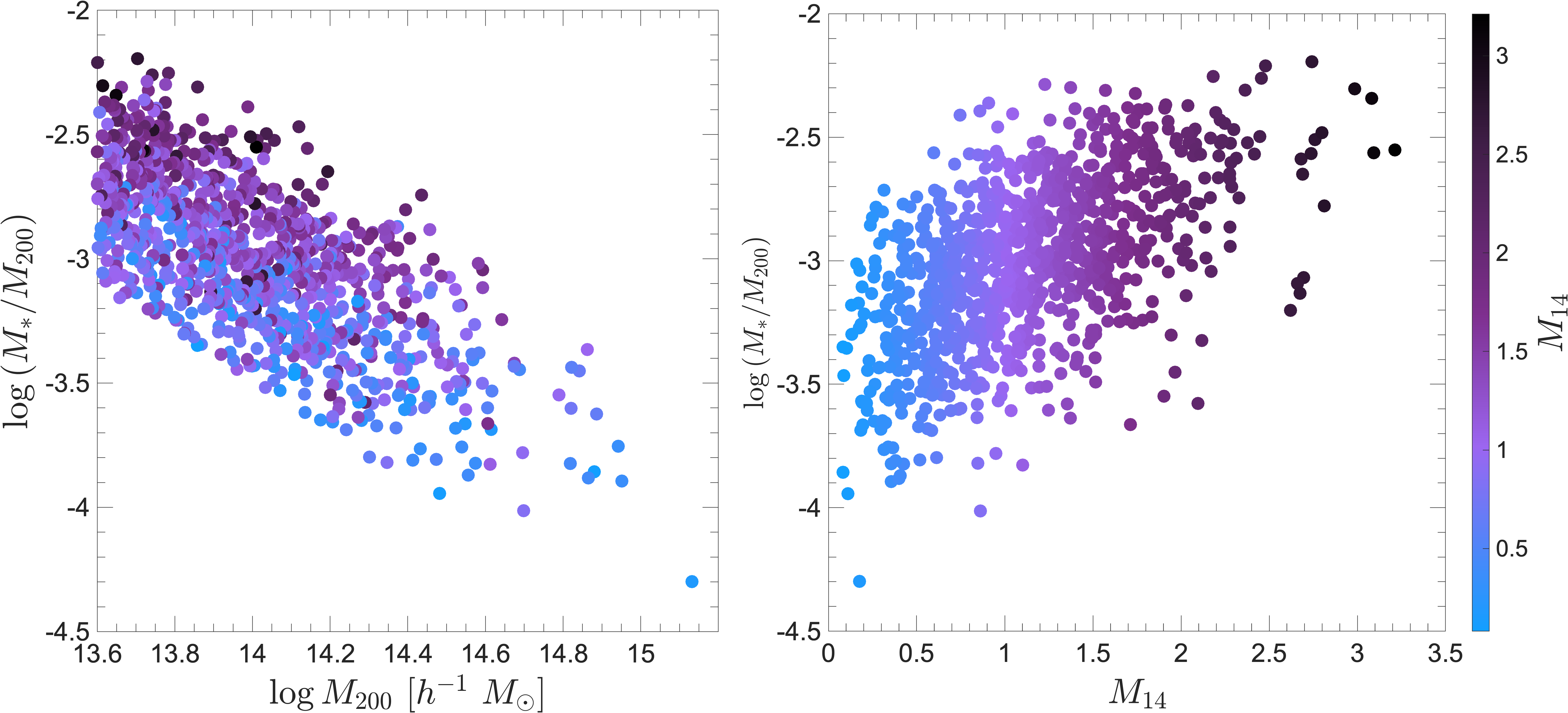} \vspace{-.5cm}
    \caption{
    Left: Distribution of the stellar-to-halo mass ratio, $\log{(M_\ast/M_{200})}$, as a function of cluster mass, $\log{M_{200}}$. 
    Right: Distribution of $\log{(M_\ast/M_{200})}$ versus the magnitude gap $M_{14}$ between the BCG and the fourth brightest galaxy member. 
    In both panels, the color bar indicates the value of $M_{14}$.
    }
   \label{fig:MhMgap}
\end{figure*}
\section{Data and Methodology} \label{sec:data}

In this section, we briefly describe the $\mathtt{GalWCat19}$ galaxy cluster catalog\footnote{\url{http://cdsarc.u-strasbg.fr/viz-bin/cat/J/ApJS/246/2}}, originally introduced by 
\citet{Abdullah20a}, 
which provides the foundation for our analysis of the BCG stellar mass–halo mass (SMHM) relation. We then outline the methodology used to fit the $\mathrm{SMHM}$ relations and to determine the best-fit parameters: the normalization ($\alpha$), slope ($\beta$), and intrinsic scatter ($\sigma_\mathrm{int}$).

\subsection{The $\mathtt{SDSS\text{-}GalWCat19}$ Galaxy Cluster Catalog} \label{sec:galw}

We use the $\mathtt{GalWCat19}$ galaxy cluster catalog, constructed from the spectroscopic data in SDSS Data Release 13 \citep{Albareti17}. Galaxies were selected based on the following criteria: (1) availability of spectroscopic observations; (2) classification as a galaxy in both the photometric and spectroscopic pipelines; (3) a spectroscopic redshift in the range $0.001 < z < 0.2$, with redshift completeness exceeding 0.7 \citep{Yang07,Tempel14}; (4) a reddening-corrected r-band magnitude $m_r< 18$; and (5) a \texttt{SpecObj.zWarning} flag of zero, indicating a reliable redshift measurement. These selection cuts resulted in a final sample of 704,200 galaxies. A full description of the catalog construction and cluster identification methodology is provided in \citet{Abdullah20a}.

Galaxy clusters in the catalog were identified using the Finger-of-God (FoG) effect \citep{Jackson72,Kaiser87,Abdullah13}, and their member galaxies were assigned using the GalWeight technique \citep{Abdullah18}. This technique operates in phase space, selecting members within a projected radius of 10~$h^{-1}$~Mpc and a line-of-sight velocity window of $\pm3500$~km~s$^{-1}$. Cluster masses were estimated using the virial mass estimator, under the assumption that mass traces the galaxy distribution \citep[e.g.,][]{Giuricin82,Merritt88}, with corrections applied for surface pressure effects to mitigate mass overestimation \citep[e.g.,][]{The86,Binney87,Abdullah11}. The virial mass was computed within the radius $R_{200}$, defined as the radius enclosing a mean overdensity of $\Delta_{200} = 200$ times the critical density of the Universe, a commonly adopted proxy for the region in hydrostatic equilibrium \citep{Carlberg97,Klypin16}.

The $\mathtt{GalWCat19}$ catalog consists of two main tables: a cluster table and a galaxy member table. The cluster table includes 1,800 clusters with redshifts in the range $0.01 < z < 0.2$ and total masses between $(0.4 - 14) \times 10^{14}~h^{-1}M_\odot$. For each cluster, the table provides sky coordinates (RA, Dec), redshift, number of member galaxies, velocity dispersion, and dynamical mass estimates computed within overdensity radii corresponding to $\Delta = 500$, 200, 100, and 5.5. Clusters identified as mergers have been excluded from the catalog.
The galaxy member table contains 34,471 galaxies located within the virial radius ($\Delta = 200$) of their host clusters. It provides each galaxy’s sky coordinates and the corresponding cluster ID.
To account for catalog incompleteness and to reduce the impact of cluster evolution and other systematic uncertainties, our analysis is restricted to the redshift range $0.02 \leq z \leq 0.125$ \citep[see][]{Abdullah20b}.

To identify the brightest cluster galaxy (BCG) in each cluster, we use the SDSS cmodel $r$-band magnitude. For all cluster members, we compute the absolute $r$-band magnitude ($M_\mathrm{r}$) using the following relation:
\begin{equation} \label{eq:AbsMag}
M_r - 5 \log {h} = m_r - DM(z) - K(z) - E(z),
\end{equation}
where $DM(z) = 5 \log D_L - 5\log h - 25$ is the distance modulus derived from the luminosity distance $D_L$. The apparent magnitude $m_r$ is converted to the AB system using the relation $m_{\mathrm{AB}} = m_{\mathrm{SDSS}} + 0.010$. The $K$-correction, $K(z)$, is calculated using version 4 of the \texttt{Kcorrect} software \citep{Blanton03a}, and the evolutionary correction, $E(z) = Q(z - 0.1)$, accounts for luminosity evolution in the $r$-band, adopting $Q = -1.62$ \citep{Blanton03a,Blanton07}. Absolute magnitudes are thus corrected to a reference redshift of $z = 0.1$, which closely matches the median redshift of the $\mathtt{GalWCat19}$ catalog ($z = 0.089$). The brightest cluster galaxy (BCG) is then identified as the member with the lowest (i.e., brightest) corrected $M_\mathrm{r}$.

The stellar masses ($M_\ast$) of the BCGs are obtained from the MPA/JHU catalog\footnote{\url{http://www.mpa-garching.mpg.de/SDSS/DR7/}}, produced by the Max Planck Institute for Astrophysics and Johns Hopkins University. These stellar masses were calculated following the methodologies described in \citet{Kauffmann03} and \citet{Salim07}. Notably, the $M_\ast$ estimates in the MPA/JHU catalog have been shown to be consistent with other independent measurements \citep[e.g.,][]{Taylor11,Leslie16}.

In Figure~\ref{fig:redshift}, we show the distribution of galaxy cluster masses (left panel) and BCG stellar masses (right panel) as functions of redshift for a sample of 1185 clusters within the range $0.02 \leq z \leq 0.125$. The color bar indicates the magnitude gap, $M_{14}$, defined as the difference in brightness between the BCG and the fourth brightest cluster member.
The left panel demonstrates that clusters with $\log{M_{200}} \geq 13.6$ [\hm] are approximately complete across the considered redshift range. The right panel displays BCGs spanning stellar masses from $9.0 < \log{M_\ast} < 12$ [\hms], with a notable decline in the number of BCGs at higher redshifts for $\log{M_\ast} \lesssim 10.5$ [\hms]. To ensure completeness and minimize bias, we define our fiducial sample by selecting BCGs with $\log{M_\ast} \geq 10.5$ [\hms]. This yields a final sample of 996 clusters (and their BCGs) within $0.02 \leq z \leq 0.125$, $\log{M_{200}} \geq 13.6$ [\hm], and $\log{M_\ast} \geq 10.5$ [\hms]. In Section~\ref{sec:sys}, we examine the impact of these selection thresholds on the best-fit parameters of the SMHM relations.

\begin{figure*}
\centering
\hspace{0 cm}  \includegraphics[width=1\linewidth]{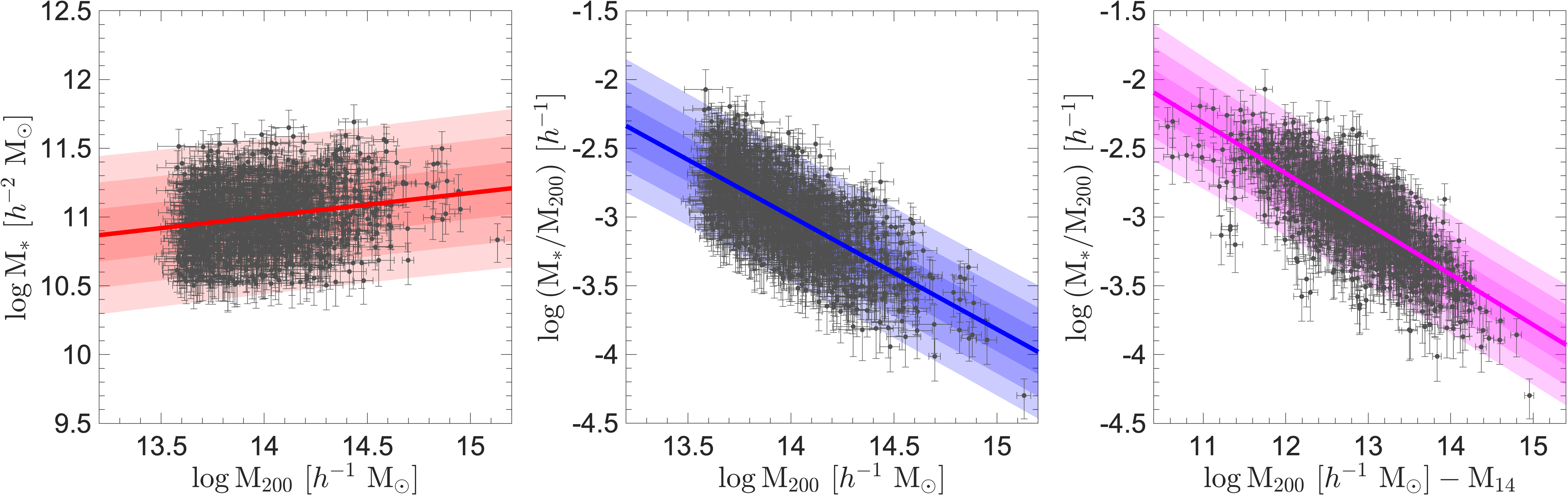} \vspace{-0.5 cm}
\caption{BCG stellar mass--halo mass relations for our cluster sample. 
Left: The standard $\mathrm{SMHM}$ relation showing $\log M_\ast$ as a function of $\log M_{200}$. 
Middle: $\mathrm{SMHM}_N$ relation showing $\log(M_\ast/M_{200})$ as a function of $\log M_{200}$, illustrating the declining stellar mass fraction with increasing halo mass.
Right: $\mathrm{SMHM}^G_N$ relation, where $\log(M_\ast/M_{200})$ is plotted against $\log{M_{200}} - M_{14}$. 
In all panels, gray points show individual clusters with $1\sigma$ error bars. Solid lines indicate the best-fit relations, and the shaded regions denote the $1\sigma$, $2\sigma$, and $3\sigma$ confidence intervals.}
\label{fig:MsMh}
\end{figure*}

In Figure~\ref{fig:MhMgap}, we show the distribution of our fiducial cluster sample in two parameter spaces: the $\log_{10}(M_\ast/M_{200})$ versus $\log_{10}(M_{200})$ plane (left) and the $\log_{10}(M_\ast/M_{200})$ versus magnitude gap $M_{14}$ plane (right).
The left panel illustrates the well-established trend in which the stellar-to-halo mass ratio decreases with increasing halo mass.
The right panel shows that $\log_{10}(M_\ast/M_{200})$ varies systematically with $M_{14}$, indicating that the magnitude gap carries additional information beyond halo mass alone. 
This motivates treating $M_{14}$ as a second parameter in the $\mathrm{SMHM}$ relation, with the goal of reducing the intrinsic scatter. 
Such a dependence is physically plausible because larger magnitude gaps are commonly associated with earlier assembly and a more evolved dynamical state, reflecting a history of BCG growth through mergers and depletion of bright satellites \citep[e.g.,][]{Raouf14,Golden-Marx18}.

\begin{table*}
\centering
\caption{
Fitting parameters for three BCG stellar mass–halo mass relations. We model $\langle \log Y|X \rangle = \alpha + \beta \log\!\left(X/X_{\mathrm{piv}}\right)$
with $X_\mathrm{piv}=14.2$.
Each row lists the best-fit normalization ($\alpha$), slope ($\beta$), and intrinsic scatter ($\sigma_{\mathrm{int}}$).
}
\label{tab:fit}
\renewcommand{\arraystretch}{1.4}
\begin{tabular}{lccc}
\hline
Relation & Normalization ($\alpha$) & Slope ($\beta$) & Intrinsic scatter ($\sigma_{\mathrm{int}}$) \\
\hline
SMHM: $Y=\log M_\ast$ vs $X=\log M_{200}$ &
11.038 $\pm$ 0.014 & 0.171 $\pm$ 0.033 & 0.188 $\pm$ 0.011 \\
$\mathrm{SMHM}_N$: $Y=\log(M_\ast/M_{200})$ vs $X=\log M_{200}$ &
-3.159 $\pm$ 0.014 & -0.823 $\pm$ 0.032 & 0.156 $\pm$ 0.013 \\
$\mathrm{SMHMG}_N$: $Y=\log(M_\ast/M_{200})$ vs $X=\log M_{200}-M_{14}$ &
-3.490 $\pm$ 0.023 & -0.366 $\pm$ 0.017 & 0.136 $\pm$ 0.013 \\
\hline
\end{tabular}
\end{table*}

Motivated by these findings, we test whether including $M_{14}$ as a secondary parameter can reduce the scatter in the stellar-to-halo mass relation. Specifically, we explore three formulations of the stellar mass–halo mass relation to investigate the connection between BCG stellar mass and cluster halo mass:
\begin{enumerate}
    \item $\log M_\ast$ vs. $\log M_{200}$: This is the standard BCG stellar mass–halo mass relation, referred to as the $\mathrm{SMHM}$ relation. It represents the baseline correlation commonly adopted in the literature.

    \item $\log(M_\ast/M_{200})$ vs. $\log M_{200}$: This is the normalized BCG stellar mass–halo mass relation, referred to as the $\mathrm{SMHM}_N$ relation. It characterizes the stellar-to-halo mass ratio as a function of halo mass and reflects stellar mass efficiency across different halo masses.

    \item $\log(M_\ast/M_{200})$ vs. $\log M_{200} - M_{14}$: This is the gap-corrected normalized BCG stellar mass–halo mass relation, referred to as the $\mathrm{SMHM}^G_N$ relation. It incorporates the magnitude gap $M_{14}$ as a secondary parameter to account for the halo assembly history and its effect on the stellar-to-halo mass ratio.
\end{enumerate}
Our goal is to evaluate whether incorporating $M_{14}$ reduces the intrinsic scatter and improves the predictability of the stellar-to-halo mass connection.
\subsection{Methodology for Fitting the Stellar–Halo Mass Relations} \label{sec:method}

We model the probability distribution of a dependent variable $Y$ at a fixed independent variable $X$ using a lognormal distribution. The probability distribution is expressed as \citep[e.g.,][]{Simet17,Chiu20}:
\begin{equation} \label{eq:prob}
\begin{split}
P(\log{Y}|X)= \frac{1}{\sqrt{2\pi\sigma^2_{\log{Y},X}}} \times ~~~~~~~~~~~~~~~~~~~~~\\
\exp{\left[-\frac{\left(\log{Y} - \left<\log{Y}|X\right>\right)^2}{2\sigma^2_{\log{Y},X}}\right]},
\end{split}
\end{equation}
\noindent where the mean relation is defined as
\begin{equation} \label{eq:rich}
\left<\log{Y}\,|\,X\right> = \alpha + \beta \log\!\left(\frac{X}{X_{\mathrm{piv}}}\right),
\end{equation}
\noindent with $X_{\mathrm{piv}}$ a fixed pivot chosen to reduce the covariance between $\alpha$ and $\beta$.
The total variance in $\log Y$ at fixed $X$ includes contributions from measurement uncertainties in both $X$ and $Y$, as well as an intrinsic scatter term,
\begin{equation} \label{eq:var}
\sigma^2_{\log{Y},X} = \beta^2\sigma^2_{\log{X}}+\sigma^2_{\log{Y}}+\sigma^2_\mathrm{int}.
\end{equation}
\noindent Here, $\alpha$ is the normalization (evaluated at $X=X_{\mathrm{piv}}$), $\beta$ is the slope of the $Y$--$X$ relation, and $\sigma_\mathrm{int}$ is the intrinsic scatter in $\log Y$ at fixed $X$.
We neglect redshift evolution since our sample spans a narrow redshift range ($0.02 \leq z \leq 0.125$), and including an explicit evolution term does not affect our results.
To estimate $\alpha$, $\beta$, and $\sigma_\mathrm{int}$, we employ the affine-invariant Markov Chain Monte Carlo (MCMC) sampler developed by \citet{Goodman10}, as implemented in the MATLAB package \texttt{GWMCMC}\footnote{\url{https://github.com/grinsted/gwmcmc}}, which is inspired by the Python package \texttt{emcee} \citep{Foreman13}.

\section{Results} 
\label{sec:results}
In this section we present the results of our analysis of the SMHM relations for brightest cluster galaxies using the \(\mathtt{GalWCat19}\) cluster sample at \(z \approx 0.1\).
We first report the best-fit relations and their uncertainties.
We then interpret the inferred SMHM slope in the high-mass regime probed by our cluster sample.
Next we quantify the intrinsic scatter and assess the improvement obtained by including the magnitude gap \(M_{14}\) as a secondary parameter.
Finally we compare our measurements with previous observational constraints and with commonly used SMHM models.

\subsection{Best-fit BCG Stellar Mass--Halo Mass Relations}
\label{sec:bestfit}

We consider three parameterizations of the stellar mass--halo mass relation introduced in Section~\ref{sec:galw}: the standard SMHM relation ($\mathrm{SMHM}$), the normalized relation ($\mathrm{SMHM}_N$), and the gap-corrected normalized relation ($\mathrm{SMHM}^G_N$), which incorporates the magnitude gap $M_{14}$ as a secondary parameter. For each relation, we derive the best-fit parameters $\alpha$, $\beta$, and $\sigma_\mathrm{int}$ by modeling the relation as
$\langle \log Y \mid X \rangle = \alpha + \beta \log\!\left(\frac{X}{X_{\mathrm{piv}}}\right)$,
with a pivot mass of $\log X_{\mathrm{piv}} = 14.2$ ($\sim$ the median cluster mass of our sample). The fitting is performed using the Markov Chain Monte Carlo procedure described in Section~\ref{sec:method}.

Figure~\ref{fig:MsMh} presents the BCG stellar mass--halo mass relations for our sample of galaxy clusters, fitted using the three models described in Section~\ref{sec:galw}. In the left panel, we show the standard $\mathrm{SMHM}$ relation, where the stellar mass of the BCG, $M_\ast$, is plotted as a function of halo mass, $M_{200}$. The middle panel shows the $\mathrm{SMHM}_N$ relation, expressed as $\log(M_\ast/M_{200})$ versus $\log M_{200}$, capturing the decreasing stellar mass fraction with increasing halo mass. The right panel shows the $\mathrm{SMHM}^G_N$ relation, where $\log(M_\ast/M_{200})$ is plotted against $\log{M_{200}} - M_{14}$, incorporating the magnitude gap as a secondary variable.
The best-fit relations are shown as red, blue, and magenta lines, with shaded regions representing the 1$\sigma$, 2$\sigma$, and 3$\sigma$ uncertainties derived from the MCMC posterior.

The best-fit parameters for each model are summarized in Table~\ref{tab:fit}. 
For the standard $\mathrm{SMHM}$ relation, we find a moderate positive slope of $\beta = 0.171$ and an intrinsic scatter of $\sigma_\mathrm{int} = 0.188$ dex. 
The $\mathrm{SMHM}_N$ relation exhibits a stronger negative slope of $\beta = -0.823$, consistent with a decreasing stellar mass fraction at higher halo masses, and a slightly reduced scatter of $0.156$ dex. Notably, the $\mathrm{SMHM}^G_N$ relation yields the lowest intrinsic scatter of $0.136$ dex, along with a moderate slope of $\beta = -0.366$. The robustness of our results to variations in the sample selection criteria is discussed in Section~\ref{sec:sys}.
We now interpret these results by examining the physical meaning of the measured slopes and the origin of the intrinsic scatter, including the role of the magnitude gap.

\subsection{Interpretation of the High-Mass SMHM Relation}

The measured slope of the standard $\mathrm{SMHM}$ relation in our cluster sample is relatively shallow, with $\beta = 0.17$, indicating that BCG stellar mass increases slowly compared to halo mass at the high-mass end. This value is consistent with slopes reported in previous observational studies of cluster-scale systems (e.g., \citealt{Gonzalez2013,Erfanianfar19,Golden-Marx19}). Such behavior is expected in halos well above the characteristic knee of the SMHM relation at $M_{\rm halo} \sim 10^{12}\,M_{\odot}$, where the efficiency of stellar mass growth begins to decline (e.g., \citealt{Behroozi13c,Moster13}). Below this mass scale, galaxies efficiently convert accreted baryons into stars, whereas above it, several physical processes act to suppress further star formation.\citep{2024MNRAS.534..361S,Pei2024}.

In the cluster regime, BCGs reside in halos where gas is shock-heated to the virial temperature and forms a hot, pressure-supported intracluster medium. The long cooling times of this gas, combined with feedback from active galactic nuclei, prevent efficient cooling and condensation onto the central galaxy (e.g., \citealt{Croton06,McNamara07,Fabian12}). As a result, in-situ star formation in BCGs is strongly suppressed at low redshift (e.g., \citealt{Donahue2010,Edwards2016}), and additional halo mass growth is dominated by dark matter accretion rather than by the formation of new stars in the BCG.

Although halo mass continues to increase through mergers and accretion, the stellar mass of the BCG grows only slowly, primarily through dry mergers with satellite galaxies (e.g., \citealt{DeLucia07,Lidman12}). Even in this merger-driven channel, a substantial fraction of the accreted stellar material is stripped during infall and deposited into the intracluster light rather than being incorporated into the BCG itself (e.g., \citealt{Purcell07,Contini14,Kravtsov18}). This inefficient transfer of stellar mass naturally leads to the shallow slope observed in the SMHM relation at the high-mass end.\citep{Golden-Marx2025}

\begin{figure*}
\centering
\hspace{0 cm}  \includegraphics[width=1\linewidth]{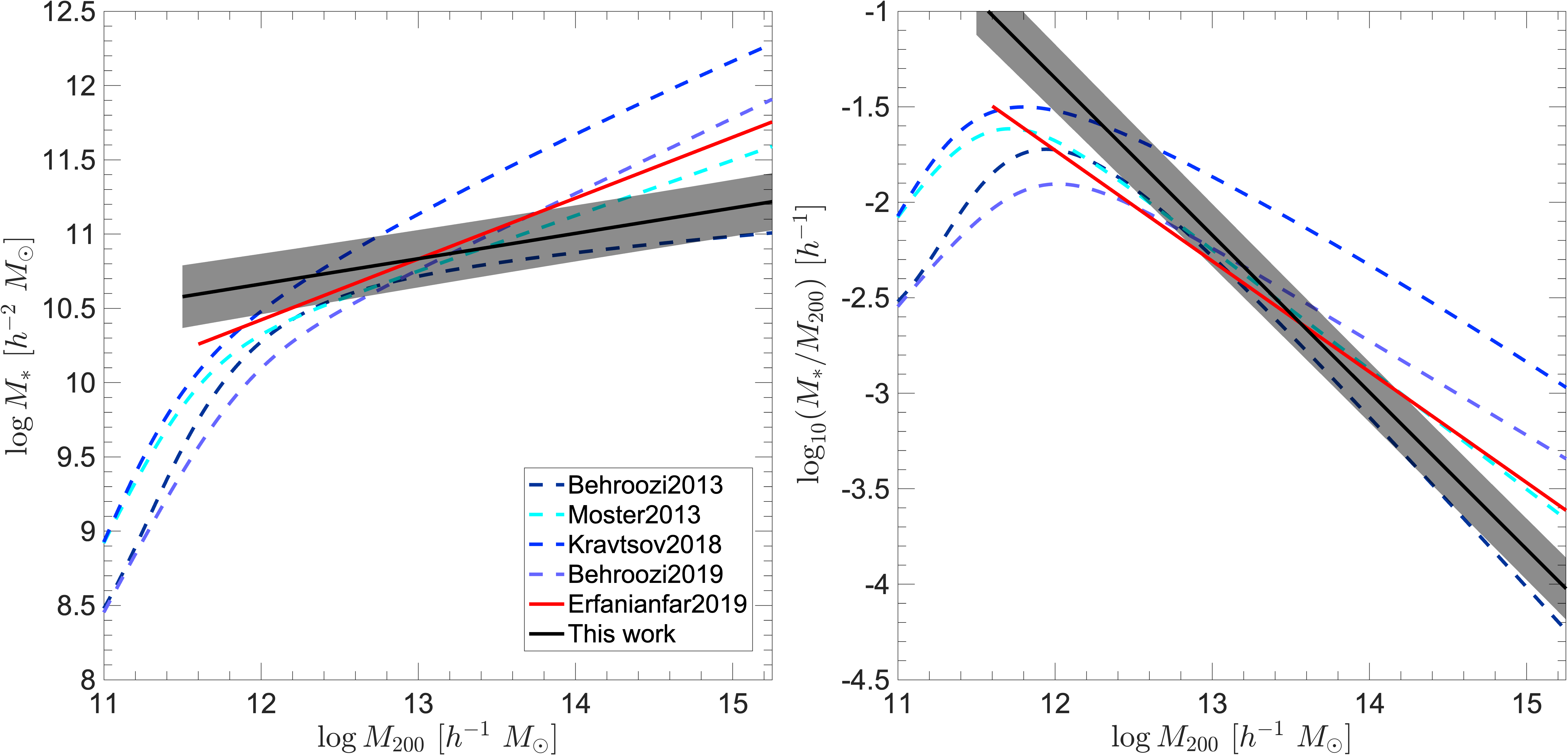} \vspace{-0.5 cm}
\caption{
Comparison between our derived BCG stellar mass-to-halo mass relations at $z \sim 0.1$ and several published models and observations. 
Left: $\log M_\ast$ versus $\log M_{200}$ relation ($\mathrm{SMHM}$). 
Right: $\log(M_\ast/M_{200})$ versus $\log M_{200}$ relation ($\mathrm{SMHM}_N$). 
The black solid line represents our best-fit relation, with the shaded region indicating the $1\sigma$ uncertainty. 
The red solid line denotes the direct observational measurement from \citet{Erfanianfar19}, while the colored dashed curves show empirical models from \citet{Behroozi13c}, \citet{Moster13}, \citet{Kravtsov18}, and \citet{Behroozi19}.
}
\label{fig:comp}
\end{figure*}

\subsection{Intrinsic Scatter and the Role of the Magnitude Gap}
\label{sec:scatter}

Despite the well-defined mean trends described above, the SMHM relations exhibit non-negligible intrinsic scatter. As shown in Figure~\ref{fig:MsMh}, clusters with comparable halo masses can host BCGs with substantially different stellar masses, indicating that halo mass alone does not fully capture the diversity of BCG growth. For the standard $\mathrm{SMHM}$ relation, the intrinsic scatter is approximately $0.19$~dex. This scatter likely arises from differences in halo assembly history, merger activity, and dynamical state, which affect the efficiency with which stellar mass is accumulated (e.g., \citealt{Zhang16,Golden-Marx22}). In the absence of parameters that encode the relevant physical information, the SMHM relation therefore remains intrinsically broad.

Including the magnitude gap, $M_{14}$, as a secondary parameter leads to a clear reduction in the intrinsic scatter. For the gap-corrected normalized relation, $\mathrm{SMHM}^G_N$, the intrinsic scatter decreases by approximately $30\%$, reaching $\sigma_\mathrm{int} = 0.136$~dex. This reduction indicates that the magnitude gap captures information beyond halo mass alone and helps organize a fraction of the observed dispersion in BCG stellar mass at fixed halo mass. In this sense, $M_{14}$ acts as a proxy for aspects of the cluster assembly history that influence the relative dominance of the BCG.

Physically, the effectiveness of $M_{14}$ reflects its connection to the formation time and merger history of the cluster. At fixed halo mass, systems with large magnitude gaps are typically more evolved, having experienced earlier assembly and, on average, more efficient merging of their most massive satellites into the central galaxy, resulting in a more dominant BCG (e.g., \citealt{Dariush10}). In contrast, clusters with small magnitude gaps tend to be less dynamically mature, with stellar mass still distributed among several bright satellites and signs of ongoing accretion activity (e.g., \citealt{Raouf14}). The effectiveness of $M_{14}$ therefore arises from its sensitivity to cluster assembly history rather than halo mass itself, consistent with the substantial scatter observed in $M_{14}$ at fixed $M_{200}$.

The remaining scatter in the $\mathrm{SMHM}^G_N$ relation likely reflects a combination of residual physical variation and observational effects, including uncertainties in BCG stellar mass estimates, halo mass measurements, and projection or selection effects, as well as additional physical processes not explicitly captured by the present parameterization. Overall, these results demonstrate that a two-parameter description of the SMHM relation provides a more complete representation of the data than a one-parameter model based solely on halo mass.

\subsection{Comparison with Previous Observations and Models}
We compare our results with previous observational measurements and widely used  models at $z \sim 0.1$. Figure~\ref{fig:comp} shows this comparison for both the standard SMHM relation (left panel) and the normalized SMHM relation (right panel). 
First, we compare our results with the  observational relation of \citet{Erfanianfar19} (red line). \citet{Erfanianfar19} find a steeper SMHM scaling in the left panel ($\beta=0.41$) than our best-fit relation ($\beta=0.17$), and correspondingly a shallower decline in the normalized relation in the right panel ($\beta=-0.58$) than ours ($\beta=-0.82$). 
The slope offset is plausibly driven by methodological differences between the two analyses. 
Our relations are derived from a large, homogeneous, spectroscopically confirmed cluster sample (996 systems), with BCG stellar masses taken from the MPA/JHU catalog, cluster masses derived self-consistently from the same spectroscopic catalog, and the fit explicitly including an intrinsic scatter term via MCMC. Moreover, our analysis is restricted to low redshift (0.02 \(\le z \le\) 0.125), so evolutionary effects are expected to be minimal. For the \citet{Erfanianfar19}  analysis, they used X-ray selected systems and derived BCG stellar masses from SED fitting to catalog photometry (SDSS cModel combined with GALEX and WISE), and their BCG identification relies on photometric information (with spectroscopy when available). Note that neither our analysis nor \citet{Erfanianfar19} includes the intracluster light  component in the BCG stellar mass.

We next compare our relations with widely used models at $z \sim 0.1$, including \citet{Behroozi13c}, \citet{Moster13}, \citet{Kravtsov18}, and \citet{Behroozi19}. Figure~\ref{fig:comp} exhibits substantial variation among these models, particularly at the high-mass end, underscoring the level of systematic uncertainty inherent in deriving the SMHM relation. Although all models broadly reproduce the characteristic double power-law behavior, they differ significantly in both normalization and slope, especially for $\log M_{200} \gtrsim 12~[h^{-1}M_\odot]$. These discrepancies arise from differences in observational inputs, mass definitions, and modeling assumptions. 
First, the adopted definition of stellar mass plays a major role: some models consider only the stellar mass of the central galaxy, while others include the extended envelope and ICL. Including the ICL increases the normalization at the high-mass end, as in the \citet{Kravtsov18} relation. 
Second, photometric systematics (surface-brightness limits, profile extrapolation, and sky subtraction) can shift the inferred stellar masses of massive galaxies and BCGs at the $\sim 0.2$ dex level \citep[e.g.,][]{Bernardi13,Bernardi17,Li22}. 
Third, choices of stellar population synthesis (SPS) models, star-formation histories (SFH), and the IMF can change the stellar-mass scale by $\sim 0.2$--0.3 dex, producing vertical offsets among models \citep[e.g.,][]{Conroy09}. 
Finally, variations in halo-mass definitions ($M_{200c}$, $M_{200m}$, or $M_{\mathrm{vir}}$) and the assumed concentration--mass relation used to convert between them can introduce shifts in the halo-mass scale at the $\sim 0.1$ dex level (and can approach $\sim 0.2$ dex depending on mass and redshift) \citep[e.g.,][]{Hu03,Diemer15}.

In summary, stellar mass estimates are model dependent, and systematic offsets are expected across pipelines owing to differences in data and assumptions (e.g., spectroscopy versus photometry-based SED fitting). Such effects are most pronounced for massive galaxies, where $\sim 0.2$ dex shifts are common. We find that SDSS Granada photometry-based SED masses are higher than MPA/JHU by $\sim 0.2$ dex at the massive end.

Cluster mass estimation is similarly challenging. Cluster masses can be inferred from galaxy dynamics (e.g., virial mass estimators), weak gravitational lensing, or X-ray observations, but each approach is subject to method-specific systematics. Dynamical masses can be biased by projection effects, interlopers, velocity anisotropy, departures from equilibrium, and substructure or nearby large-scale structure. X-ray based masses can be biased if the intracluster medium deviates from hydrostatic equilibrium, while lensing masses can be affected by line-of-sight structure and projection. These effects can lead to biased mass estimates and increased scatter in scaling relations (see, e.g., \citealp{Tonry81,The86,Fadda96,Abdullah13,Zhang19}).

\section{Conclusion}\label{sec:conc}

In this work we use the spectroscopic \(\mathtt{GalWCat19}\) catalog to investigate the stellar-to-halo mass (SMHM) relation of brightest cluster galaxies (BCGs) at low redshift (\(0.02 \le z \le 0.125\)). The spectroscopic nature of the catalog, together with the FoG-GalWeight toolkit, provides robust membership assignments and self-consistent dynamical mass estimates, thereby mitigating projection effects and minimizing the need for evolutionary corrections.

We examine three formulations of the SMHM relation: the standard relation (\(M_{\star}\)--\(M_{200}\)), the normalized relation (\(M_{\star}/M_{200}\)--\(M_{200}\)), and a gap-corrected normalized relation that includes the magnitude gap \(M_{14}\). While the standard and normalized relations recover the expected trends, they exhibit non-negligible intrinsic scatter. Incorporating \(M_{14}\) as a secondary parameter yields the tightest correlation, reducing the intrinsic scatter to \(\sigma_{\mathrm{int}} = 0.136~\mathrm{dex}\). This result indicates that the magnitude gap carries additional information about cluster assembly that is not captured by halo mass alone, and improves the predictive power of the SMHM relation.

We also quantify the impact of selection and systematic effects by varying richness, redshift, cluster-mass, and BCG stellar-mass thresholds. In all cases the resulting shifts in normalization, slope, and intrinsic scatter remain modest and are smaller than the statistical uncertainties, demonstrating that our conclusions are robust to reasonable changes in sample definition.

Overall our findings support the use of BCG stellar mass as a tracer of cluster-scale dark matter halos, while highlighting the importance of secondary parameters such as \(M_{14}\) for capturing the imprint of assembly history. The stability of the relation to selection choices, combined with the reduced scatter achieved when including \(M_{14}\), provides useful constraints on the co-evolution of BCGs and their host clusters and informs models of hierarchical structure formation and massive-galaxy growth in dense environments.

\section*{Acknowledgments}
GW gratefully acknowledges support from the National Science Foundation through grant AST-2347348.

\renewcommand{\thesection}{Appendix~\Alph{section}} 
\renewcommand{\thesubsection}{\Alph{section}.\arabic{subsection}} 
\renewcommand{\thefigure}{A\arabic{figure}}     
\renewcommand{\thetable}{A\arabic{table}}       
\renewcommand{\theequation}{A\arabic{equation}} 

\setcounter{section}{0} 
\setcounter{figure}{0}
\setcounter{table}{0}
\setcounter{equation}{0}

\begin{table*}
\centering
\caption{
Systematic uncertainties in the normalization ($\alpha$), slope ($\beta$), and intrinsic scatter ($\sigma_\mathrm{int}$) arising from variations in sample selection criteria, including richness, redshift, cluster mass, and BCG stellar mass thresholds. Each $\Delta$ value represents the standard deviation of the corresponding parameter across the tested threshold range, quantifying its sensitivity to the selection cuts. The three SMHM formulations refer to: the standard relation (SMHM), the normalized relation (SMHM$_N$), and the gap-corrected normalized relation (SMHM$^G_N$).
}
\label{tab:sys}
\begin{tabular}{lccc}
\hline
Relation & $\Delta\alpha$  & $\Delta\beta$ & $\Delta\sigma_{\mathrm{int}}$  \\
\hline
& \multicolumn{3}{c}{Varying richness threshold} \\
\hline
$\mathrm{SMHM}$: $\log M_\ast - \log M_{200}$ & 0.004 & 0.023 & 0.001 \\
$\mathrm{SMHM}_N$: $\log (M_\ast/M_{200}) - \log M_{200}$ & 0.003 & 0.021 & 0.005 \\
$\mathrm{SMHM}^G_N$: $\log (M_\ast/M_{200}) - (\log M_{200} - M_{14})$ & 0.008 & 0.014 & 0.003 \\
\hline
& \multicolumn{3}{c}{Varying lower redshift threshold} \\
\hline
$\mathrm{SMHM}$: $\log M_\ast - \log M_{200}$ & 0.006 & 0.004 & 0.001 \\
$\mathrm{SMHM}_N$: $\log (M_\ast/M_{200}) - \log M_{200}$ & 0.006 & 0.004 & 0.001 \\
$\mathrm{SMHM}^G_N$: $\log (M_\ast/M_{200}) - (\log M_{200} - M_{14})$ & 0.004 & 0.007 & 0.005 \\
\hline
& \multicolumn{3}{c}{Varying upper redshift threshold} \\
\hline
$\mathrm{SMHM}$: $\log M_\ast - \log M_{200}$ & 0.020 & 0.016 & 0.002 \\
$\mathrm{SMHM}_N$: $\log (M_\ast/M_{200}) - \log M_{200}$ & 0.020 & 0.016 & 0.002 \\
$\mathrm{SMHM}^G_N$: $\log (M_\ast/M_{200}) - (\log M_{200} - M_{14})$ & 0.012 & 0.005 & 0.009 \\
\hline
& \multicolumn{3}{c}{Varying cluster mass threshold} \\
\hline
$\mathrm{SMHM}$: $\log M_\ast - \log M_{200}$ & 0.002 & 0.025 & 0.002 \\
$\mathrm{SMHM}_N$: $\log (M_\ast/M_{200}) - \log M_{200}$ & 0.003 & 0.024 & 0.004 \\
$\mathrm{SMHM}^G_N$: $\log (M_\ast/M_{200}) - (\log M_{200} - M_{14})$ & 0.022 & 0.017 & 0.013 \\
\hline
& \multicolumn{3}{c}{Varying BCG stellar mass threshold} \\
\hline
$\mathrm{SMHM}$: $\log M_\ast - \log M_{200}$ & 0.048 & 0.035 & 0.048 \\
$\mathrm{SMHM}_N$: $\log (M_\ast/M_{200}) - \log M_{200}$ & 0.049 & 0.031 & 0.060 \\
$\mathrm{SMHM}^G_N$: $\log (M_\ast/M_{200}) - (\log M_{200} - M_{14})$ & 0.043 & 0.009 & 0.029 \\
\hline
\end{tabular}
\end{table*}

\begin{figure*}
\centering
\hspace{0 cm}    \includegraphics[width=0.8\linewidth]{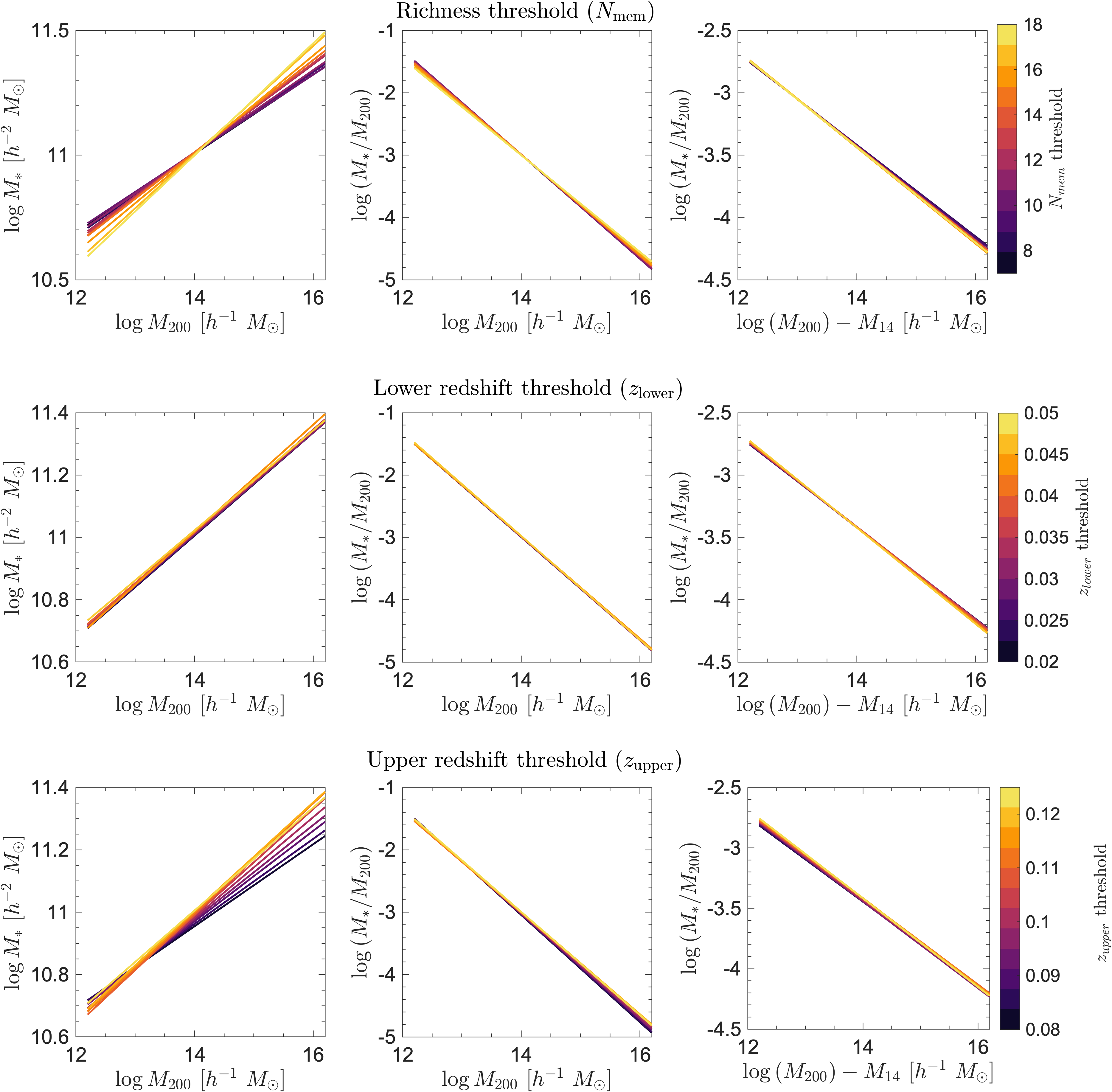} \vspace{-0.5 cm}
\caption{
Impact of varying selection thresholds on the SMHM relations. 
Each row shows the best-fit SMHM (left), SMHM$_N$ (middle), and SMHM$^G_N$ (right) relations for different cuts in richness ($N_{\mathrm{mem}}$, top row), lower redshift ($z_{\mathrm{lower}}$, middle row), and upper redshift ($z_{\mathrm{upper}}$, bottom row). 
Color scales indicate the threshold values used. 
The relations remain nearly unchanged across the tested ranges, demonstrating that systematic effects from these selection criteria are small.}
\label{fig:sys1}
\end{figure*}

\section{Systematic Uncertainties} \label{sec:sys}
In this section, we assess the systematic uncertainties associated with our sample selection criteria and their impact on the derived SMHM relation parameters. Specifically, we investigate how variations in five key thresholds influence the best-fit values of $\alpha$, $\beta$, and $\sigma_\mathrm{int}$:
\begin{enumerate}
    \item Richness threshold: We vary the minimum number of cluster members ($N_{\mathrm{mem}}$) used to define a cluster from 8 to 18 in steps of 1.
    \item Lower redshift threshold ($z_{\mathrm{lower}}$): We vary the minimum redshift from 0.020 to 0.050 in steps of 0.005.
    \item Upper redshift threshold ($z_{\mathrm{upper}}$): We vary the maximum redshift from 0.080 to 0.125 in steps of 0.005.
    \item Cluster mass threshold: We vary the minimum halo mass from $\log M_{200} = 13.6$ to $14.1$ [\hm] in steps of 0.05. 
    \item BCG stellar mass threshold: We vary the minimum stellar mass from $\log M_\ast = 10.0$ to $10.9$ [\hms] in steps of 0.05.
\end{enumerate}

\begin{figure*}
\centering
\hspace{0 cm}    \includegraphics[width=0.8\linewidth]{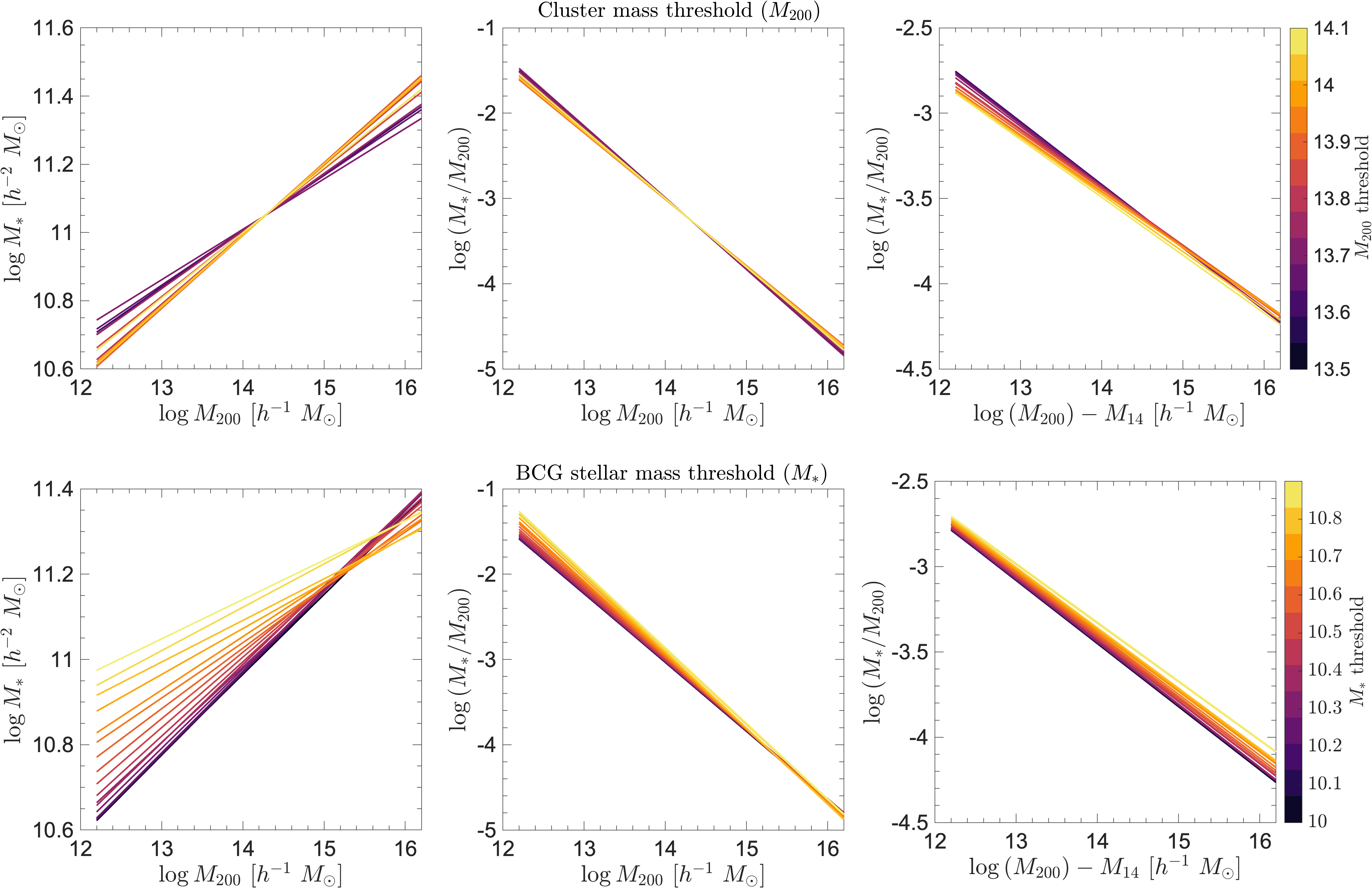} \vspace{-0.5 cm}
\caption{
Same as Figure \ref{fig:sys1}.
The top row shows the best-fit SMHM (left), SMHM$_N$ (middle), and SMHM$^G_N$ (right) relations for different cuts in cluster mass ($M_{200}$), while the bottom row shows the corresponding results for different BCG stellar mass thresholds ($M_\ast$). 
The relations remain stable across varying $M_{200}$ thresholds, while the $M_\ast$ threshold introduces somewhat larger deviations, though still modest, confirming the overall robustness of the results.}
\label{fig:sys2}
\end{figure*}

As discussed in Section~\ref{sec:galw}, our fiducial cluster (BCG) sample is defined by $N_{\mathrm{mem}} \geq 8$, $\log M_{200} \geq 13.6$ [\hm], $\log M_\ast \geq 10.5$ [\hms], and $0.02 \leq z \leq 0.125$. For each threshold variation, we re-fit the SMHM, SMHM$_N$, and SMHM$^G_N$ relations using the same MCMC procedure described in Section~\ref{sec:method}. We quantify the sensitivity of each relation by computing the standard deviation of the best-fit parameters ($\alpha$, $\beta$, and $\sigma_{\mathrm{int}}$) across the tested threshold range. These systematic uncertainties are summarized in Table~\ref{tab:sys}, where each $\Delta$ value represents the standard deviation of the corresponding parameter. Figures~\ref{fig:sys2} and \ref{fig:sys1} show the corresponding best-fit relations for the different cuts, which remain nearly unchanged across the tested ranges.

Across all tested selection thresholds the systematic uncertainties in the normalization ($\Delta\alpha$), slope ($\Delta\beta$), and intrinsic scatter ($\Delta\sigma_{\mathrm{int}}$) remain small. Variations in richness, redshift, and cluster mass thresholds indicate a very weak dependence of the SMHM parameters on these cuts. The largest shifts appear for the BCG stellar mass threshold, but even here the deviations are modest ($\leq 0.06$), confirming that the results are robust. Overall, the systematic effects are minor compared to the statistical uncertainties, demonstrating the stability of the SMHM relations against reasonable variations in sample selection.

\bibliography{Ref}{}
\bibliographystyle{aasjournal}
\appendix

\end{document}